\documentclass[3p,times,procedia]{elsarticle}
\usepackage{nupha_ecrc}

\volume{00}

\firstpage{1}

\journalname{Nuclear Physics A}

\runauth{}

\jid{nupha}

\jnltitlelogo{Nuclear Physics A}

\usepackage{amssymb}

\usepackage[figuresright]{rotating}

\newcommand {\hic}	{{\sc hic}}
\newcommand {\rhic}	{{\sc rhic}}
\newcommand {\lhc}	{{\sc lhc}}
\newcommand {\dft}	{{\sc dft}}
\newcommand {\ampt}	{{\sc ampt}}

\newcommand {\mcg}	{{\sc mcg}}
\newcommand {\ws}	{{\sc ws}}
\newcommand {\cme}	{{\mbox{\sc cme}}}
\newcommand {\cs}	{{\sc cs}}

\newcommand {\RP}	{{\sc rp}}
\newcommand {\PP}	{{\sc pp}}
\newcommand {\EP}	{{\sc ep}}

\newcommand {\psiRP}	{\psi_{_{\rm RP}}}
\newcommand {\psiPP}	{\psi_{_{\rm PP}}}
\newcommand {\psiEP}	{\psi_{_{\rm EP}}}

\newcommand {\etwo}	{\epsilon_2}
\newcommand {\epsi}	{\etwo\{\psi\}}
\newcommand {\eRP}	{\etwo\{\psiRP\}}
\newcommand {\ePP}	{\etwo\{\psiPP\}}
\newcommand {\vv}	{v_2}
\newcommand {\vRP}	{\vv\{\psiRP\}}
\newcommand {\vEP}	{\vv\{\psiEP\}}

\newcommand {\rbf}	{{\bf r}}

\newcommand {\vpsi}	{\vv\{\psi\}}
\newcommand {\Au}	{$^{197}_{\;\,79}$Au}
\newcommand {\Cu}	{$^{62}_{29}$Cu}
\newcommand {\Pb}	{$^{207}_{\;\,82}$Pb}
\newcommand {\Ru}	{$^{96}_{44}$Ru}
\newcommand {\Zr}	{$^{96}_{40}$Zr}
\newcommand {\CuCu}	{CuCu}
\newcommand {\PbPb}	{PbPb}
\newcommand {\RuRu}	{RuRu}
\newcommand {\ZrZr}	{ZrZr}
\newcommand {\AuAu}	{AuAu}
\newcommand {\bkg}	{{\mbox{\sc bkg}}}

\newcommand {\zdc}	{{\sc zdc}}
\newcommand {\rcme}	{r}

\newcommand {\psiPPEP}	{\psi_{_{\rm PP(EP)}}}
\newcommand {\psione}	{\psi_{_{1}}}

\newcommand {\av}	{a_{\vv}^{\rm EP}}
\newcommand {\aB}	{a_{_{\Bsq}}}

\newcommand {\aBEP}	{\aB^{\rm EP}}
\newcommand {\RPP}	{R^{\rm PP}}
\newcommand {\REP}	{R^{\rm EP}}

\newcommand {\rEP}	{R_{\rm EP}}

\newcommand {\RPPEP}	{R^{\rm PP(EP)}}

\newcommand {\Bbf}	{\mathbf{B}}
\newcommand {\Bsq}	{B_{\rm sq}}
\newcommand {\Bpsi}	{\Bsq\{\psi\}}
\newcommand {\BRP}	{\Bsq\{\psiRP\}}
\newcommand {\BPP}	{\Bsq\{\psiPP\}}

\newcommand {\psitwo}	{\psi_{_{2}}}
\newcommand {\gOS}	{\gamma_{_{\rm OS}}}
\newcommand {\gSS}	{\gamma_{_{\rm SS}}}
\newcommand {\dg}	{\Delta\gamma}
\newcommand {\dgpsi}	{\dg\{\psi\}}

\newcommand {\mean}[1]	{\langle #1\rangle}

\begin{document}

\begin{frontmatter}



\dochead{XXVIIth International Conference on Ultrarelativistic Nucleus-Nucleus Collisions\\ (Quark Matter 2018)}

\title{Re-examining the premise of isobaric collisions and a novel method to measure the chiral magnetic effect}

\author[label1]{Hao-jie Xu}
\author[label2]{Jie Zhao}
\author[label1]{Xiaobao Wang}
\author[label3]{Hanlin Li}
\author[label4,label5]{Zi-Wei Lin}
\author[label1]{Caiwan Shen}
\author[label1,label2]{Fuqiang Wang}
\address[label1]{School of Science, Huzhou University, Huzhou, Zhejiang 313000, China}
\address[label2]{Department of Physics and Astronomy, Purdue University, West Lafayette, Indiana 47907, USA}
\address[label3]{College of Science, Wuhan University of Science and Technology, Wuhan, Hubei 430065, China}
\address[label4]{Department of Physics, East Carolina University, Greenville, North Carolina 27858, USA}
\address[label5]{Key Laboratory of Quarks and Lepton Physics (MOE) and Institute of Particle Physics, Central China Normal University, Wuhan, Hubei 430079, China}

\begin{abstract}
In this proceeding we will show that the expectations of the isobaric $^{96}_{44}\mathrm{Ru}+^{96}_{44}\mathrm{Ru}$ and $^{96}_{40}\mathrm{Zr}+^{96}_{40}\mathrm{Zr}$ collisions on chiral magnetic effect (CME) search may not hold as originally anticipated due to large uncertainties in the isobaric nuclear structures. We demonstrate this using Woods-Saxon densities and the proton and neutron densities calculated by the density functional theory. Furthermore, a novel method is proposed to gauge background and possible CME contributions in the same system, intrinsically better than the isobaric collisions of two different systems.  We illustrate the method with Monte Carlo Glauber and AMPT (A Multi-Phase Transport) simulations.

\end{abstract}


\end{frontmatter}


\section{Introduction}
\label{sec:introduction}

In quantum chromodynamics (QCD), the interactions of quarks with topological gluon fields can induce chirality imbalance and parity violation in local domain under the approximate chiral symmetry restoration~\cite{Kharzeev:1998kz}. A chirality imbalance could lead to an electric current, or charge separation (\cs) 
in the direction of a strong magnetic field ($\Bbf$). 
This phenomenon is called the chiral magnetic effect (\cme)~\cite{Kharzeev:1998kz}. 
Searching for the \cme\ is one of the most active research in heavy ion collisions (\hic)~\cite{Kharzeev:2015znc,Zhao:2018ixy}. In \hic\ the \cs\ is commonly measured by the three-point correlator~\cite{Voloshin:2004vk}, $\gamma\equiv\cos(\phi_\alpha+\phi_\beta-2\psiRP)$, where $\phi_\alpha$ and $\phi_\beta$ are the azimuthal angles of two charged particles, and $\psiRP$ is that of the reaction plane (\RP, spanned by the impact parameter and beam directions) to which the $\Bbf$ produced by the incoming protons is perpendicular on average. 
Positive $\dg\equiv\gOS-\gSS$ ({\sc os}:opposite-sign, {\sc ss}:same -sign) signals, consistent with the \cme-induced \cs\ perpendicular to the \RP, have been observed~\cite{Adamczyk:2014mzf}. 
The signals are, however, inconclusive because of a large charge-dependent background arising from particle correlations (e.g.~resonance decays) coupled with the elliptic flow anisotropy ($\vv$)~\cite{Wang:2009kd}. 

To better control the background, isobaric collisions of \Ru+\Ru\ (\RuRu) and \Zr+\Zr\ (\ZrZr) have been proposed~\cite{Voloshin:2010ut}. One expects their backgrounds to be almost equal because of the same mass number, while the atomic numbers, hence $\Bbf$, differ by 10\%. 
This is verified by {\em Monte Carlo} Glauber (\mcg) calculations using the Woods-Saxon (\ws) density profile~\cite{Deng:2016knn}.
As a net result, the \cme\ signal-to-background ratio would be improved by over a factor of 7 in comparative measurements between RuRu and ZrZr collisions than in each of them individually~\cite{Deng:2016knn}.  
The isobaric collisions are planned for 2018 at \rhic; they would yield a \cme\ signal of $5\sigma$ significance with the projected data volume, 
if one assumes that the \cme\ contributes 1/3 of the current $\dg$ measurement in \AuAu\ collisions.

However, there can be non-negligible deviations of the Ru and Zr nuclear densities from \ws. In this proceeding, we will show their effects on the sensitivity of isobaric collisions for the \cme\ search~\cite{Xu:2017zcn}, and a novel method will be proposed to avoid those uncertainties~\cite{Xu:2017qfs}.  

\section{Re-examining the premise of isobaric collisions}
\begin{figure}[hbt]
	  \begin{minipage}[t]{0.35\linewidth}
  \begin{center}
   	 \includegraphics[width=1.0\textwidth]{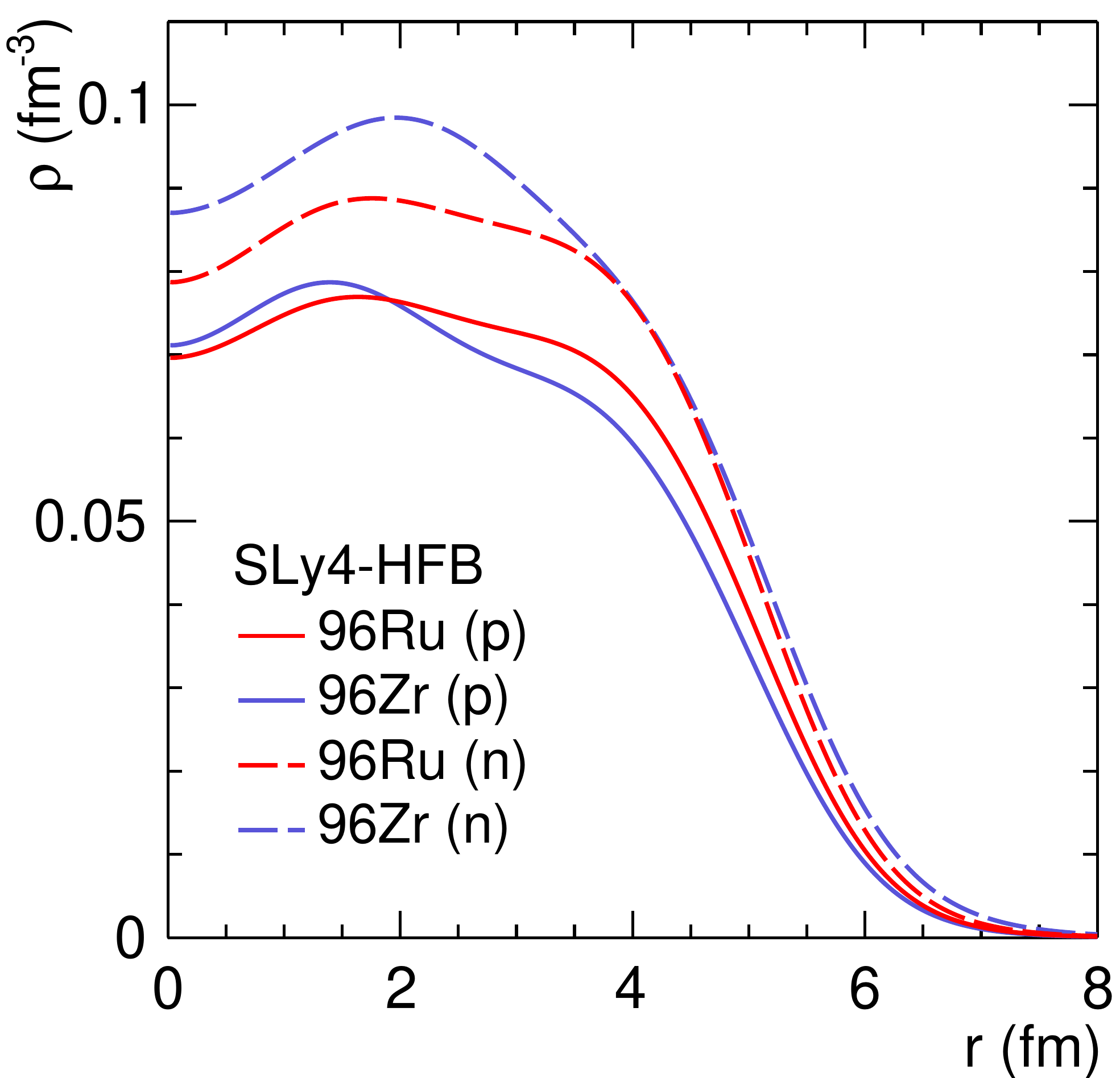}
  \end{center}
  \vspace{-0.2in}
  \caption{(Color online) Proton and neutron density distributions of the \Ru\ and \Zr\ nuclei, assumed spherical, calculated by the \dft\ method.}
  \label{fig:rho}
	  \end{minipage}
	  \hspace*{0.08\textwidth}
	  \begin{minipage}[t]{0.55\linewidth}
  \begin{center}
    \includegraphics[width=1.0\textwidth]{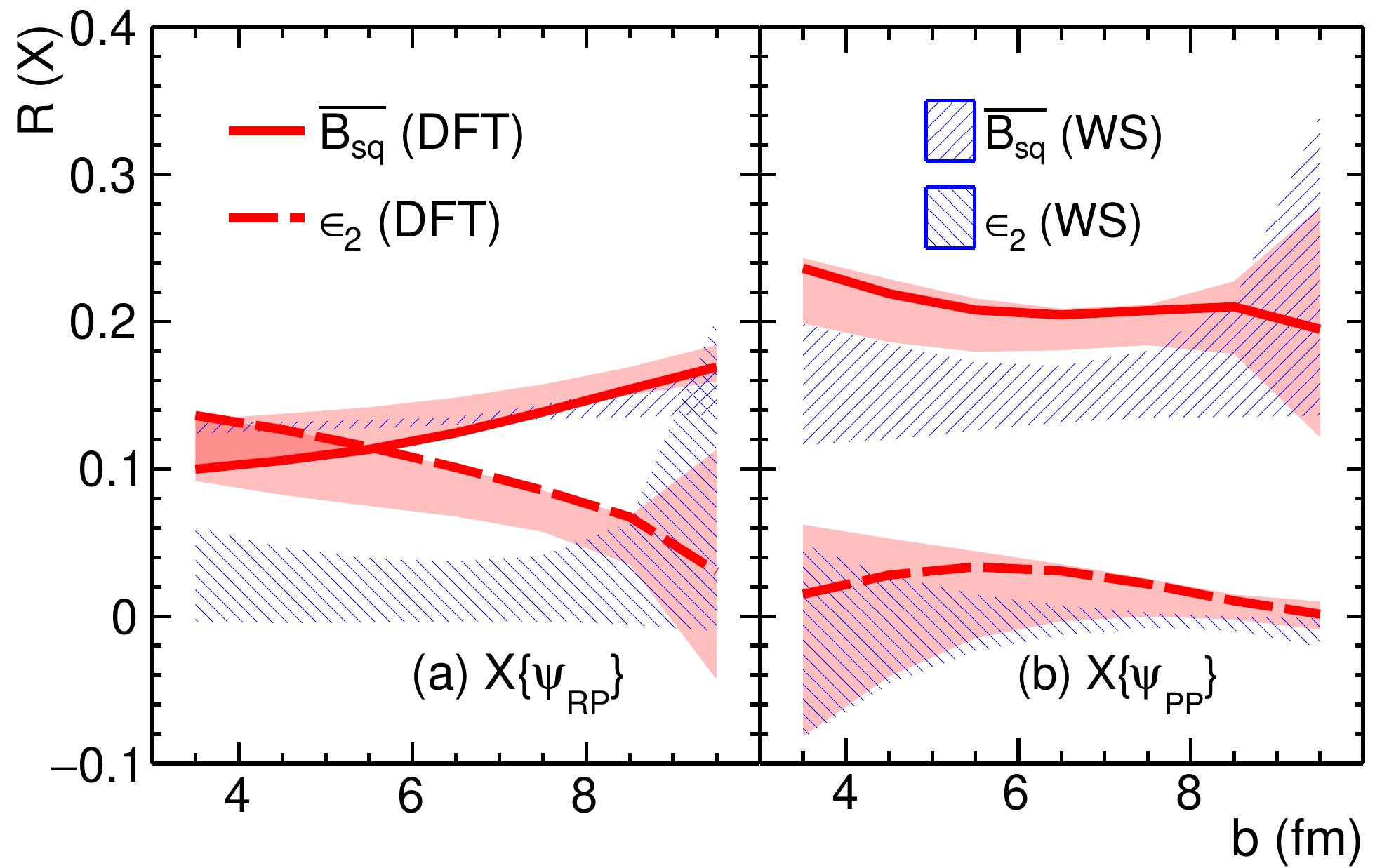}
  \end{center}
  \vspace{-0.2in}
  \caption{(Color online) Relative differences between \RuRu\ and \ZrZr\ collisions in $\epsi$ and $\Bpsi$ with respect to (a) $\psi=\psiRP$ and (b) $\psi=\psiPP$, using the \dft\ densities. The shaded areas correspond to \dft\ density uncertainties from Ru and Zr deformities; the hatched areas show the corresponding results using \ws\ density distributions.
}
  \label{fig:R}
	  \end{minipage}
\end{figure}

Because of the different numbers of protons--which suffer from Coulomb repulsion--and neutrons, the structures of the \Ru\ and \Zr\ nuclei must not be identical.
By using density functional theory (\dft) , we calculate the Ru and Zr proton and neutron distributions using the well-known SLy4 mean field including pairing correlations (Hartree-Fock-Bogoliubov, HFB approach)~\cite{Wang:2016rqh}.
Those density distributions are shown in Fig.~\ref{fig:rho}.
Protons in Zr are more concentrated in the core, while protons in Ru, 10\% more than in Zr, are pushed more toward outer regions. The neutrons in Zr, four more than in Ru, are more concentrated in the core but also more populated on the nuclear skin. Theoretical uncertainties are estimated by using different sets of density functionals, SLy5 and SkM* for the mean field, with and without pairing (HFB/HF), and found to be small. 

The $\etwo$ of the transverse overlap geometry in \RuRu\ and \ZrZr\ collisions is calculated event-by-event with \mcg~\cite{Xu:2014ada}, using the \dft\ nucleon densities in Fig.~\ref{fig:rho}.
$\Bbf(\rbf,t=0)$ is also calculated 
for \RuRu\ and \ZrZr\ collisions using the \dft\ proton densities. The calculations follow Ref.~\cite{Deng:2012pc}, with a finite proton radius (0.88~fm~ is used but the numeric value is not critical) to avoid the singularity at zero relative distance.  
Two reference planes are used for each collision system: reaction plane ($\psiRP$) and participant plane ($\psiPP$).
For the \cme\ search with isobaric collisions, the relative differences in $\etwo$ and $\Bsq$ are of importance. Figure~\ref{fig:R} shows the relative differences $R(\ePP)$, $R(\eRP)$, $R(\BPP)$, and $R(\BRP)$; $R(X)$ is defined as
\begin{equation}
R(X)\equiv2(X_{\rm\RuRu}-X_{\rm\ZrZr})/(X_{\rm\RuRu}+X_{\rm\ZrZr})\;
\end{equation}
where $X_{\rm\RuRu}$ and $X_{\rm\ZrZr}$ are the $X$ values in \RuRu\ and \ZrZr\ collisions, respectively. The thick solid curves are the default results with the \dft\ densities in Fig.~\ref{fig:rho} . 
The shaded areas correspond to theoretical uncertainties bracketed by the two \dft\ density cases where Ru is deformed with $\beta_2=0.158$ and Zr is spherical and where Ru is spherical and Zr is deformed with $\beta_2=0.217$. 
The hatched areas represent our results using \ws\ densities with the above two cases of nuclear deformities.

We also investigate whether our density profiles would, in a dynamical model, lead to a final-state $\vv$ difference between \RuRu\ and \ZrZr\ collisions and whether the $\Bsq$ difference preserves with respect to the event plane (\EP) reconstructed from the final-state particle momenta. 
We employ A Multi-Phase Transport (\ampt) model with ``string melting''~\cite{Lin:2004en}, 
which can reasonably reproduce heavy ion bulk data at \rhic\ and the \lhc~. We found that the general trends are similar to those in Fig.~\ref{fig:R}~\cite{Xu:2017zcn}.

From the \mcg\ and \ampt\ simulations, we find that the \dft\ nuclear densities, together with the Woods-Saxon (\ws) densities, yield wide ranges of differences in $\Bsq$ with respect to the participant plane (\PP) and the reaction plane (\RP). 
It is further found that those nuclear densities introduce, in contrast to \ws, comparable differences in $\eRP$ ($\vRP$) and $\BRP$ with respect to the reaction plane (\RP), deminishing the premise of isobaric collisions to help identify the \cme. With respect to the participant plane (\PP), the $\ePP$ ($\vEP$) difference can still be sizable, as large as $\sim3$\%, possibly weakening the power of isobaric collisions for the \cme\ search~\cite{Xu:2017zcn}. 

\section{A movel method to measure the \cme}
\begin{figure*}
  \begin{center}
    \includegraphics[width=0.83\textwidth]{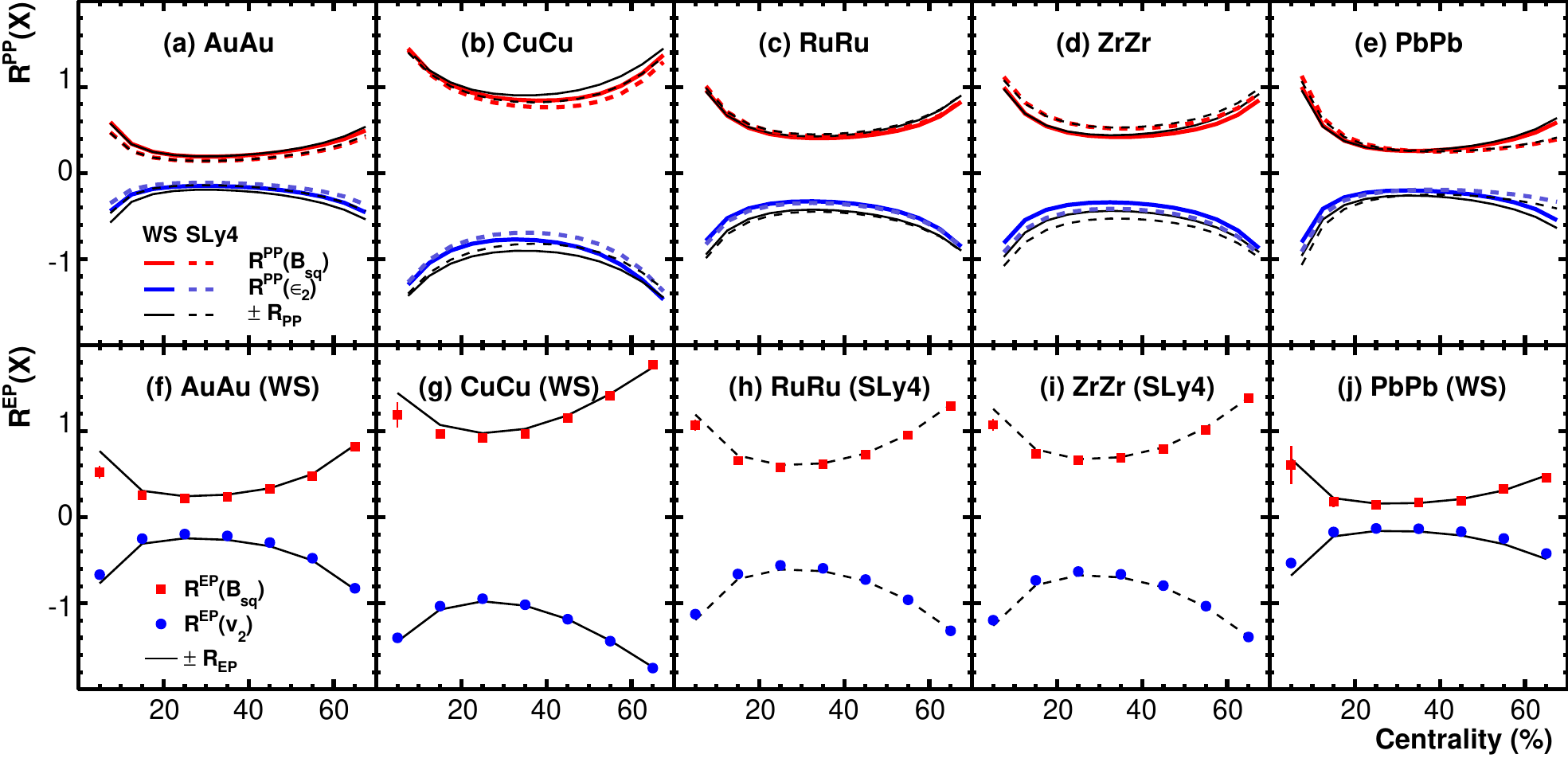}
  \vspace{-0.2in}
  \caption{\label{fig}(Color online) Relative differences $\RPP(\etwo)$, $\RPP(\Bsq)$, $\RPP$ from \mcg\ (upper panel) and $\REP(\vv)$, $\REP(\Bsq)$, $\rEP$ from \ampt\ (lower panel) for (a,f) \AuAu, (b,g) \CuCu, (c,h) \RuRu, and (d,i) \ZrZr\ at \rhic, and (e,j) \PbPb\ at the \lhc. Both the \ws\ and \dft-calculated densities are shown for the \mcg\ results, while the used density profiles are noted for the \ampt\ results. Errors, mostly smaller than the symbol size, are statistical.}
  \end{center}
\end{figure*}

Based on the above study~\cite{Xu:2017zcn}, we found that with respect to $\psiPP$, $\vv$ is stronger than that with respect to $\psiRP$. This is because elliptic flow ($\vv$) develops in relativistic heavy ion collisions from the anisotropic overlap geometry of the participant nucleons.
The magnetic field ($\Bbf$) is, on the other hand, produced mainly by spectator protons and its direction fluctuates nominally about $\psiRP$, not $\psiPP$. Therefore, $\Bbf$ with respect to $\psiPP$ is weaker than $\Bbf$ with respect to $\psiRP$. 
Our new method is based on the opposite behaviors in the fluctuations of the magnetic field and $v_2$ in a single nucleus-nucleus collision, thus bears minimal theoretical and experimental uncertainties.
It is convenient to define a relative difference~\cite{Xu:2017qfs},
\begin{equation}
\RPPEP(X)\equiv2\cdot\frac{X\{\psiRP\}-X\{\psiPPEP\}}{X\{\psiRP\}+X\{\psiPPEP\}}\;,
\end{equation}
where $X\{\psiRP\}$ and $X\{\psiPPEP\}$ are the measurements of quantity $X$ with respect to $\psiRP$ and $\psiPP$ (or $\psiEP$ described below), respectively. 
The upper panels of Fig.~\ref{fig} show $\RPP(\etwo)$ and $\RPP(\Bsq)$ calculated by a \mcg\ model for \Au+\Au\ (\AuAu), \Cu+\Cu\ (\CuCu), \RuRu,
\ZrZr\ collisions at \rhic\ and \Pb+\Pb\ (\PbPb) collisions at the \lhc. 
Although a theoretical concept, the \RP\ may be assessed by Zero-Degree Calorimeters (\zdc) measuring sidewards-kicked spectator neutrons (directed flow $v_1$)~.
The lower panels of Fig.~\ref{fig} show \ampt\ simulation results of $\REP(\vv)$ and $\REP(\Bsq)$, compared to $\pm\rEP$ ($\rEP \equiv 2(1 - \mean{\cos2(\psiEP-\psiRP)}/\mathcal{R}_{EP})/(1 + \mean{\cos2(\psiEP-\psiRP)}/\mathcal{R}_{EP})$, where $\mathcal{R}_{EP}$ is the \EP\ resolution). 

The commonly used $\dg$ variable contains, in addition to the \cme\ it is designed for, $\vv$-induced background, 
$\dgpsi=\cme(\Bpsi)+\bkg(\vpsi)$.
$\dgpsi$ can be measured with respect to $\psi=\psiRP$ (using the 1st order event plane $\psione$ by the \zdc) and $\psi=\psiEP$ (2nd order event plane $\psitwo$ via final-state particles). 
If $\bkg(\vv)$ is proportional to $\vv$~ and $\cme(\Bsq)$ to $\Bsq$~, then 
\begin{equation}
\REP(\dg)=2\frac{r(1-\aBEP)-(1-\av)}{r(1+\aBEP)+(1+\av)}\approx\frac{1-r}{1+r}\REP(\vv)\;.
\end{equation}
Here $\rcme\equiv\cme(\BRP)/\bkg(\vEP)$ can be considered as 
the relative \cme\ signal to background contribution.
If the experimental measurement $\REP(\dg)$ equals to $\REP(\vv)$ (i.e.~$\dg$ scales like $\vv$), then \cme\ contribution is zero; if $\REP(\dg)\approx-\REP(\vv)$ (i.e.~$\dg$ scales like $\Bsq$), then background is close to zero and all would be \cme; and if $R(\dg)=0$, then background and \cme\ contributions are of similar magnitudes.
Recently, our new method has been  applied to experimental data by the STAR collaboration, see Ref.~\cite{Zhao:2018en} for more details.

\section{Summary} 
To reduce background effects in \cme\ search, isobaric \Ru+\Ru\ and \Zr+\Zr\ collisions have been proposed where the $\vv$-induced backgrounds are expected to be similar while the \cme-induced signals to be different. In our  study, the proton and neutron density distributions of \Ru\ and \Zr\ are calculated using the energy density functional theory (\dft). They are then implemented in the {\em Monte Carlo} Glauber (\mcg) model to calculate the eccentricities ($\etwo$) and magnetic fields ($\Bbf$), and in A Multi-Phase Transport (\ampt) model to simulate the $\vv$. 
It is found that those nuclear densities, together with the Woods-Saxon (\ws) densities, yield wide ranges of differences in $\Bsq$ with respect to the participant plane (\PP) and the reaction plane (\RP). 

We thus propose a novel method with comparative measurements of $\dg$ with respect to $\psiRP$ and $\psiPP$ in the same collision system. Our method is superior to isobaric collisions where large systematics persist. The novel method has been applied to experimental data by the STAR collaboration.
With improved statistics, the novel method we report here should be able to decisively answer the question of the CME in quantum chromodynamics.

This work was supported in part by the National Natural Science Foundation of China under Grants No.~11647306, 11747312, U1732138, 11505056, 11605054, and 11628508, and US~Department of Energy under Grant No.~DE-SC0012910.





\begin{thebibliography}{10}
\expandafter\ifx\csname url\endcsname\relax
  \def\url#1{\texttt{#1}}\fi
\expandafter\ifx\csname urlprefix\endcsname\relax\def\urlprefix{URL }\fi
\expandafter\ifx\csname href\endcsname\relax
  \def\href#1#2{#2} \def\path#1{#1}\fi

\bibitem{Kharzeev:1998kz}
D.~Kharzeev, R.~D. Pisarski, M.~H.~G. Tytgat, Phys. Rev. Lett. 81 (1998) 512--515.

\bibitem{Kharzeev:2015znc}
D.~E.~Kharzeev, J.~Liao, S.~A.~Voloshin and G.~Wang, Prog. Part. Nucl. Phys.  88 (2016) 1.

\bibitem{Zhao:2018ixy}
J.~Zhao, Int. J. Mod. Phys. A33~(13) (2018) 1830010.

\bibitem{Voloshin:2004vk}
S.~A. Voloshin, Phys. Rev. C70
  (2004) 057901.

\bibitem{Adamczyk:2014mzf}
L.~Adamczyk, et~al., Phys. Rev. Lett. 113 (2014)
  052302.

\bibitem{Wang:2009kd}
F.~Wang, Phys. Rev. C81 (2010) 064902.
S.~Pratt, S.~Schlichting and S.~Gavin, Phys. Rev. C84 (2011) 024909.
A.~Bzdak, V.~Koch and J.~Liao, Phys. Rev. C83 (2011) 014905.
F.~Wang and J.~Zhao, Phys. Rev. C95 (2017) 051901.

\bibitem{Voloshin:2010ut}
S.~A. Voloshin, Phys. Rev. Lett. 105 (2010) 172301.

\bibitem{Deng:2016knn}
W.-T. Deng, X.-G. Huang, G.-L. Ma, G.~Wang, Phys. Rev. C94 (2016) 041901.

\bibitem{Xu:2017zcn}
H.-j. Xu, X.~Wang, H.~Li, J.~Zhao, Z.-W. Lin, C.~Shen, F.~Wang, Phys.
  Rev. Lett. 121 (2018) 022301.

\bibitem{Xu:2017qfs}
H.-j. Xu, J.~Zhao, X.~Wang, H.~Li, Z.-W. Lin, C.~Shen, F.~Wang, Chin. Phys. C42 (2018) 084103.

\bibitem{Wang:2016rqh}
X.~B. Wang, J.~L. Friar, A.~C. Hayes, Phys. Rev. C94~(3) (2016) 034314.

\bibitem{Xu:2014ada}
H.-j. Xu, L.~Pang, Q.~Wang, Phys. Rev. C89~(6) (2014) 064902.

\bibitem{Deng:2012pc}
W.-T. Deng, X.-G. Huang, Phys. Rev. C85 (2012) 044907.

\bibitem{Lin:2004en}
Z.-W. Lin, C.~M. Ko, B.-A. Li, B.~Zhang, S.~Pal, Phys. Rev. C72 (2005) 064901.

\bibitem{Zhao:2018en}
J.~Zhao, {Quark Matter 2018 proceedings}, arXiv:1807.09925.

\end{thebibliography}







\end{document}